**Detuning-dependent narrowing of Mollow triplet lines of driven quantum dots**

A P Saiko[1], R Fedaruk[2] and S A Markevich[1]

[1]Scientific-Practical Materials Research Centre NAS of Belarus, Minsk, Belarus

[2] Institute of Physics, University of Szczecin, 70-451, Szczecin, Poland

E-mail: saiko@ifttp.bas-net.by; fedaruk@wmf.univ.szczecin.pl

We study the two-time correlation function and the resonance fluorescence spectrum of a semiconductor quantum dot excited by a strong off-resonant laser pulse. The obtained analytical expressions exhibit a specific detuning-dependent damping of Rabi oscillations of the dressed quantum dot as well as a detuning-dependent width of Mollow-triplet lines. In the absence of pure dephasing, the central peak of the triplet is broadened, upon increasing detuning, but the blue and red side peaks are narrowed. We demonstrate that pure dephasing processes can invert these dependences. A crossover between the regimes of detuning-dependent narrowing and broadening of the side and central peaks is identified. The predicted effects are consistent with resent experimental results and numerical calculations.

PACS number(s): 42.50.Hz, 78.67.Hc, 81.07.Ta, 78.47.jp

Emission properties of driven semiconductor quantum dots (QDs) have attracted much interest recently due to their potential applications in the fields of photonics and quantum information technology [1-3]. Semiconductor QDs provide an atomlike light-matter interaction demonstrating typical quantum dynamical features of isolated natural atoms. In particular, artificial atoms as such QDs excited by strong resonant continuous-wave optical field exhibit resonance fluorescence spectrum containing three peaks, known as the Mollow triplet [4]. Under pulsed resonant excitation in time-resolved experiments, QDs undergo the Rabi oscillations [5,6]. In the dressed-atom approach [7], these effects are understood as resulting from the quantum transitions in the total coupled system of atom and driving photons. In contrast with natural atoms, QDs interact with their solid-state environment in a more complicated manner. There are a variety of loss mechanisms for quantum dots [5]. The damping of the Rabi oscillations as well as the width of the Mollow-triplet lines can be used for identifying these mechanisms. One of the main consequences of the solid-state character of QDs is specific dephasing caused by coupling to acoustic phonons. A well-known signature of phonon coupling is revealed in an excitation-induced dephasing (EID) with a rate proportional to the square of the effective Rabi frequency. Experimental evidence of EID effects has been recently observed as oscillation damping in pulsed photocurrent measurements on a resonantly driven QD [5]. The effect of EID has also been observed as the Mollow-triplet sideband broadening under resonant continuous-wave excitation of a single QD in a microcavity [1]. In that paper, the phenomenon of the spectral Mollow sideband *narrowing* in dependence of laser-excitation detuning from the bare emitter resonance has been demonstrated as well, but that effect had to be left open for further theoretical analysis. Previous numerical studies using the polaron master equation (ME) approach with cavity coupling did not uncover the narrowing effect [8]. Recently, it has been shown that a crossover between the regimes of detuning-dependent sideband narrowing and broadening can be qualitatively understood from a theoretical model based on the polaron ME without inclusion of the QED-cavity coupling [3]. Numerical calculations presented in paper [9] have also shown that for off-resonant driving, narrowing in the spectral sideband width can occur in certain conditions. In paper [10] it has been shown that the decay rate of the Rabi oscillations in QDs can decrease with increasing detuning.

In the present paper, we analytically calculate the two-time correlation function and resonance fluorescence spectrum of a semiconductor QD excited by an off-resonant laser pulse taking into account the exciton-phonon interaction. Our analytical results allow us to give the clear physical treatment of the detuning-dependent narrowing and broadening of the Mollow-triplet lines. We show that pure dephasing processes radically influence the character of these phenomena and can cause a crossover between the detuning-dependent narrowing and broadening of the triplet peaks.

We model the QD as a two-level system with an energy splitting $\omega_0$ between ground $|1\rangle$ and excited $|2\rangle$ states (we take $\hbar = 1$). The QD is driven by a coherent laser field of frequency $\omega_L$ and is coupled to a phonon bath. A master equation for the density matrix $\rho$ of the QD [3] in the Markovian approximation can be written as:

$$i\frac{\partial \rho}{\partial t} = [H, \rho] + i\Lambda\rho, \quad (1)$$

where

$$H = \Delta s^z + \frac{\Omega}{2}(s^+ + s^-) \quad (2)$$

is the Hamiltonian of the QD in the frame rotating at frequency $\omega_L$, and

$$\Lambda\rho = \frac{\gamma_{21} + \gamma_{ph}^-}{2} D[s^-]\rho + \frac{\gamma_{12} + \gamma_{ph}^+}{2} D[s^+]\rho + \frac{\eta}{2} D[s^z]\rho - \gamma_{ph}^{cd}(s^+\rho s^+ + s^-\rho s^-) \quad (3)$$

is the relaxation superoperator. Here $s^{\pm,z}$ are components of the pseudospin operator, describing the QD state and satisfying the commutation relations: $[s^+, s^-] = 2s^z$, $[s^z, s^\pm] = \pm s^\pm$, $\Delta = \omega_0 - \omega_L$, $\Omega = \Omega_0 e^{-W(T)/2}$, where $\Omega_0$ is the bare Rabi frequency that describes the coherent exciton pumping from the laser field, and $e^{-W(T)}$ is the Debye-Waller factor, which takes into account the effect of acoustic phonons on the coherent laser-QD interaction and depends on a bath temperature $T$ and the electron-phonon coupling. This factor describes the so-called elastic processes in which the momentum obtained by the QD from the exciting photon is transmitted to a whole crystal without a phonon emission or absorption. In addition, $D[O]\rho = 2O\rho O^+ - O^+O\rho - \rho O^+O$, $\gamma_{21}$ and $\gamma_{12}$ are the rates of photon radiative processes from the excited state $|2\rangle$ of the QD to its ground state $|1\rangle$ and vice versa, the $\gamma_{ph}^-$ process corresponds to enhanced radiative decay, while the $\gamma_{ph}^+$ process represents an incoherent excitation process caused by the exciton-phonon interaction (non-elastic photon-QD scattering processes are realized through multi-phonon transitions), $\gamma_{ph}^{cd}$ is the cross-dephasing rate induced by phonons, and $\eta$ is the dephasing rate introduced phenomenologically. As it was shown in Ref. [3], the relaxation rate $\gamma_{ph} = \gamma_{ph}^- + \gamma_{ph}^+$ weakly depends on $\Delta$ in the detuning range considered below and can be approximated as $\gamma_{ph} = k\Omega^2$ where $k$ is a coefficient. For sufficiently large driven strengths, $\Omega \gg \gamma_{ph}^{cd}$, the cross-dephasing term in Eq. (3) can be neglected. This term rapidly oscillates in the interaction representation with the Hamiltonian (2) and gives only corrections of the second and higher orders in $\gamma_{ph}^{cd}/\Omega$.

After the canonical transformation $\rho_1 = u^+ \rho u$, where $u = e^{-\theta(s^+ - s^-)/2}$, the ME (1) is transformed into

$$i\frac{\partial \rho_1}{\partial t} = [H_1, \rho_1] + i\Lambda_1 \rho_1, \qquad (4)$$

$H_1 = u^+ H u - iu^+ \frac{\partial u}{\partial t} = \varepsilon s^z$, $\Lambda_1 = u^+ \Lambda u = \frac{\Gamma_\downarrow}{2} D[s^-] + \frac{\Gamma_\uparrow}{2} D[s^+] + \frac{\Gamma_\varphi}{2} D[s^z]$,

where $\varepsilon = (\Delta^2 + \Omega^2)^{1/2}$,

$$\Gamma_\downarrow = \frac{1}{4}(\gamma_{21} + \gamma_{12} + \gamma_{ph}^- + \gamma_{ph}^+)(1 + \cos^2\theta) + \frac{1}{2}(\gamma_{21} + \gamma_{ph}^- - \gamma_{12} - \gamma_{ph}^+)\cos\theta + \frac{1}{4}\eta \sin^2\theta,$$

$$\Gamma_\uparrow = \frac{1}{4}(\gamma_{21} + \gamma_{12} + \gamma_{ph}^- + \gamma_{ph}^+)(1 + \cos^2\theta) - \frac{1}{2}(\gamma_{21} + \gamma_{ph}^- - \gamma_{12} - \gamma_{ph}^+)\cos\theta + \frac{1}{4}\eta \sin^2\theta,$$

$\Gamma_\varphi = \eta \cos^2\theta + (\gamma_{21} + \gamma_{12} + \gamma_{ph}^- + \gamma_{ph}^+)\sin^2\theta$; $\cos\theta = \Delta/\varepsilon$, and $\sin\theta = \Omega/\varepsilon$.

We obtain the relaxation superoperator in the rotating wave approximation (RWA). In this case, since at the strong laser-QD interaction $\Gamma_\downarrow, \Gamma_\uparrow, \Gamma_\varphi \ll \varepsilon$, the non-diagonal terms that contain the products of spin operator pairs $s^\pm$ and $s^z$, $s^+$ and $s^+$, $s^-$ and $s^-$ are neglected in the structure of the operator.

The solution of equation (4) can be written as follows:

$$\rho_1(t) = e^{(-iL_1 + \Lambda_1)t}\rho_1(0). \tag{5}$$

The superoperator $L_1$ acts in an accordance with the rule: $L_1 X = [H_1, X]$. The density matrix $\rho(t)$ in the laboratory frame is given by

$$\rho(t) = u_1 \rho_1(t) u_1^+, \tag{6}$$

where $\rho_1(t)$ is defined by equation (4) and $\rho_1(0) = u_1^+ \rho(0) u_1$. Moreover, if the QD is in the ground state, $\rho(0) = 1/2 - s^z$, and we have $\rho_1(0) = 1/2 + (1/2)(s^+ + s^-)\sin\theta - s^z\cos\theta$. One can check by a direct calculation that the following relations are fulfilled:

$e^{(-iL_1 + \Lambda_1)t} s^\pm = e^{(\mp i\varepsilon - \Gamma_\perp)t} s^\pm$, $e^{(-iL_1 + \Lambda_1)t} s^z = e^{-\Gamma_\parallel t} s^z$,

$$e^{(-iL_1 + \Lambda_1)t} a = [1 + 2\sigma_0(1 - e^{-\Gamma_\parallel t}) s^z] a, \tag{7}$$

where $a = const$, $\sigma_0 = -(\Gamma_\downarrow - \Gamma_\uparrow)/\Gamma_\parallel = -(\gamma_{21} + \gamma_{ph}^- - \gamma_{12} - \gamma_{ph}^+)\cos\theta/\Gamma_\parallel$, \hfill (8)

$\Gamma_\parallel = \Gamma_\downarrow + \Gamma_\uparrow = \tilde\gamma_\parallel + (\tilde\gamma_\perp - \tilde\gamma_\parallel)\sin^2\theta$, $\Gamma_\perp = (\Gamma_\downarrow + \Gamma_\uparrow + \Gamma_\varphi)/2 = \tilde\gamma_\perp - (1/2)(\tilde\gamma_\perp - \tilde\gamma_\parallel)\sin^2\theta$,

$\tilde\gamma_\parallel = \gamma_\parallel + \gamma_{ph}$, $\tilde\gamma_\perp = \gamma_\perp + \gamma_{ph}/2$, $\gamma_\parallel = \gamma_{12} + \gamma_{21}$, $\gamma_\perp = (\gamma_{12} + \gamma_{21} + \eta)/2$, $\gamma_{ph} = \gamma_{ph}^- + \gamma_{ph}^+$.

Using Eqs. (4) - (8), we obtain the density matrix in the laboratory frame

$$\rho_{lab}(t) = \frac{1}{2} + \frac{1}{4}\left\{\left[\sin\theta(\cos\theta + 1)e^{-i(\omega_L + \varepsilon)t - \Gamma_\perp t} + \sin\theta(\cos\theta - 1)e^{-i(\omega_L - \varepsilon)t - \Gamma_\perp t}\right]s^+ + H.c.\right\} -$$

$$-\frac{1}{2}\sin\theta\left[\cos\theta e^{-\Gamma_\parallel t} + \frac{\Gamma_\downarrow - \Gamma_\uparrow}{\Gamma_\parallel}(1 - e^{-\Gamma_\parallel t})\right](s^+ e^{-i\omega_L t} + H.c.) -$$

$$-\left\{e^{-\Gamma_\perp t}\sin^2\theta\cos\varepsilon t + \cos\theta\left[\cos\theta e^{-\Gamma_\parallel t} + \frac{\Gamma_\downarrow - \Gamma_\uparrow}{\Gamma_\parallel}(1 - e^{-\Gamma_\parallel t})\right]\right\}s^z. \tag{9}$$

We now find the two-time correlation function

$g^{(1)}(t) = <s^-(t)s^+(0)> = <s^-(t)>_{s^+(0)\rho(0)} =$

$$= \frac{1}{2}\sin^2\theta e^{-i\omega_L t - \Gamma_\perp t} + \frac{1}{4}(1 + \cos\theta)^2 e^{-i(\varepsilon + \omega_L)t - \Gamma_\perp t} + \frac{1}{4}(1 - \cos\theta)^2 e^{i(\varepsilon - \omega_L)t - \Gamma_\perp t} \tag{10}$$

and the spectral density of the emitted radiation

$$S(\omega) = \frac{1}{\pi} \text{Re} \int_0^\infty dt e^{i\omega t} <s^-(t)s^+(0)> =$$

$$= \frac{1}{2\pi} \left[ \frac{\Gamma_\| \sin^2 \theta}{\Gamma_\|^2 + (\omega - \omega_L)^2} + \frac{\Gamma_\perp (1 + \cos \theta)^2 / 2}{\Gamma_\perp^2 + (\omega - \varepsilon - \omega_L)^2} + \frac{\Gamma_\perp (1 - \cos \theta)^2 / 2}{\Gamma_\perp^2 + (\omega + \varepsilon - \omega_L)^2} \right]. \quad (11)$$

It follows from Eqs. (8) and (10) that the respective decay rates of the oscillations at the side $\omega_L \pm \varepsilon$ and central $\omega_L$ frequencies are

$$\Gamma_\perp = \frac{1}{2}(\gamma_\| + \gamma_{ph} + \eta) + \frac{1}{4}(\gamma_\| + \gamma_{ph} - \eta)\frac{\Omega^2}{\Delta^2 + \Omega^2}, \quad \Gamma_\| = \gamma_\| + \gamma_{ph} - \frac{1}{2}(\gamma_\| + \gamma_{ph} - \eta)\frac{\Omega^2}{\Delta^2 + \Omega^2}. \quad (12)$$

According to Eq. (11), the full-width at half-maximum (FWHM) of the side and central lines of the Mollow triplet are equal to $2\Gamma_\perp$ and $2\Gamma_\|$, respectively.

We conclude from Eqs. (8) and (12) that the rates of the transverse $\tilde{\gamma}_\perp$ and longitudinal $\tilde{\gamma}_\|$ relaxations determine the character of detuning dependences of the decay rates for the oscillations at the side $\omega_L \pm \varepsilon$ and central $\omega_L$ frequencies. If $\tilde{\gamma}_\perp > \tilde{\gamma}_\|$, for fixed driven strength, the function $\sin^2 \theta = \Omega^2 / (\Delta^2 + \Omega^2)$ decreases with increasing detuning $\Delta$, which results in increasing the decay rate $\Gamma_\perp$ of the side oscillations. At the same time, the decay rate $\Gamma_\|$ of the central oscillation decreases. If $\tilde{\gamma}_\perp < \tilde{\gamma}_\|$, the decay rate $\Gamma_\perp$ of the side oscillations decreases with increasing detuning, while the decay rate $\Gamma_\|$ of the central oscillation increases. Obviously, the narrowing of the Mollow sidebands with increasing detuning can be realized without exciton-phonon interaction (see Eq. (12)). In this case ($\gamma_{ph} = 0$), the condition $\gamma_\perp < \gamma_\|$ must be fulfilled. This inequality can be realized when the contribution of pure dephasing processes into the linewidth is smaller than the natural linewidth. Note that at $\tilde{\gamma}_\perp = \tilde{\gamma}_\|$ (or $\gamma_\perp = \gamma_\|$ for $\gamma_{ph} = 0$), the linewidths of the sidebands as well as the central peak do not depend on detuning.

Thus, in the absence of pure dephasing, the narrowing of the Mollow sidebands with increasing laser-QD detuning is realized. Pure dephasing processes can bring about a crossover between the narrowing and broadening regime. If phonon-QD interactions do not result in pure dephasing processes and the relaxation rate $\gamma_{ph} = k\Omega^2$ is constant in the considered detuning range, phonon-QD interactions do not influence the detuning dependence and only change the condition of the crossover, $\eta = \gamma_\| + \gamma_{ph}$. Cross-relaxation processes are usually weak to change noticeably this behavior.

Fig. 1 displays the two-time correlation function and the QD emission spectrum versus the laser-QD detuning at the fixed driven strength for two dephasing rates. The correlation function is shown in the rotating frame and represents the Rabi oscillations. The off-resonant laser excitation differently modifies the intensity as well as the linewidth of each peak in the Mollow triplet. Moreover, the pure dephasing strongly influences the modification. Fig. 1 (a) and (b) displays the emission properties in the absence of pure dephasing. At $\eta = 0$, the central peak undergoes broadening and decreasing of its intensity with increasing laser-detuning. In contrast, we observe narrowing of the sidebands and different changes in intensities for the red and blue sidebands. With increasing positive (negative) detuning, the amplitude of the blue sideband increases (decreases) whereas the amplitude of the red sideband decreases (increases). Our results demonstrate a systematic narrowing of the Mollow sidebands and a broadening of the central peak with increasing laser-detuning. As one can infer from Fig. 1 (c) and (d), pure dephasing processes invert these dependences.

In Figs. 2, we plot the dependences of amplitudes of the side and central peaks on the laser detuning for spectra presented in Fig. 1. Fig. 3 depicts the FWHM values of these spectral lines. One can see that the sideband narrowing and the central-peak broadening occurs when $\eta < \gamma_{\parallel} + \gamma_{ph}$. In contrast, for $\eta > \gamma_{\parallel} + \gamma_{ph}$ the strong pure dephasing results in broadening of the sidebands and narrowing of the central peak under off-resonant excitation. Moreover, at $\eta = \gamma_{\parallel} + \gamma_{ph}$ Fig. 3 shows the crossover between the narrowing and broadening regime. In that case, the FWHM values of all peaks of the Mollow triplet are equal and do not depend on detuning.

In Fig. 4 we compare our analytical description with the experimental results for sideband broadening with increasing laser-QD detuning [3]. The parameters used are the same as in Ref. [3] except that the cross-dephasing term is neglected and the value of $\eta$ is larger (by about 10 %). Note that except for the cross-dephasing term, equations (12) coincide with the expressions in Ref. [3] for the on-resonance ($\Delta = 0$) FWHM of the side and central lines. The analytical curve is in a good agreement with the experimental data and lies between the values obtained numerically [3]. According to the theoretical prediction of Ref. [3], the FWHM values of the blue and red Mollow sidebands should have distinct broadening with increasing laser-QD detuning. The precision of experimental data is not sufficient to confirm this prediction.

As one can see from Eq. (12) and Fig. 3, the sideband narrowing occurs when $\eta < \gamma_{\parallel} + \gamma_{ph}$. In this regime, our analytical results for the sideband FWHM values are consistent with the recent numerical calculations [9] obtained at identical parameters and with the observed experimental trend [1] (Fig. 5). In a more detailed analysis of the observed narrowing effect, the systematical reducing of the Rabi frequency $\Omega$ with increasing value of $\Delta$ as well as the cavity effects should be taken into account [1]. The $\Omega$-reducing can strongly intensify the sideband narrowing effect with increasing laser-QD detuning.

In the model used, the strong laser field drastically modifies the eigenstates of a two-level QD, resulting in the additional splitting of their energy levels. Because the splitting is much smaller than the energy difference between the bare states, the appearance of low-frequency transitions (at the Rabi frequency) in the dressed QD enables the effective interaction between the QD and phonon excitations of environment. The participation of phonons in the elastic processes of the QD-photon interactions only renormalizes the constant of exciton-photon interaction. On the other hand, participating in the non-elastic processes, phonons provide an additional channel of energetic and phase relaxations. Due to the specific interaction between the QD dressed states and phonons, the relaxation rates are characterized by the quadratic dependence on the driving strength. The high-field limit used means that the Rabi frequency is much larger than the relaxation rates of the QD. Therefore, the non-RWA terms in the relaxation operator in the representation of the bare and dressed QD states can be neglected. That fact allows us to obtain the approximate analytical expressions for the time-resolved resonance fluorescence and the spectrum of scattering emission. The obtained simple explicit expressions show that the dressing of the QD states by the strong driven field mixes the relaxation rates. For example, the phase relaxation of the dressed states turns out to depend on both the phase and energetic relaxations of the bare QD. The degree of such mixing depends on the exciton-photon interaction and the laser-QD detuning.

One should note that we demonstrate the detuning-dependent sideband narrowing without cavity coupling. Moreover, this effect is not specific only for QDs and it can be observed for any two-level system when dephasing processes do not significantly broaden the natural line.

In conclusion, we have analytically described the time-resolved and spectral features of photon emission from a QD excited by an off-resonant laser pulse. We have shown that pure dephasing processes influence radically the detuning-dependent damping of the Rabi oscillations of the dressed QD as well as the width of the Mollow-triplet lines. At weak pure dephasing, the central peak of the triplet is broadened with increasing detuning, but the blue and red side peaks are narrowed. For stronger pure dephasing, the crossover between the narrowing and broadening regimes is realized. We have found a good agreement between our approximate analytical description and the recent experimental results and numerical calculations. Our approach can be applied to describe the off-resonant emission features of a wide range of two-level systems.

**References**

[1] S.M. Ulrich, S. Ates, S. Reitzenstein, A. Löffler, A. Forchel, and P. Michler, Phys. Rev. Lett. 106, 247402 (2011).

[2] C. Matthiesen, A.N. Vamivakas, and M. Atatüre, Phys. Rev. Lett. 108, 093602 (2012).

[3] A. Ulhaq, S. Weiler, S.M. Ulrich, M. Jetter, P. Michler, C. Roy, and S. Hughes, Optics Express 21, 4382 (2013).

[4] B.R. Mollow, Phys. Rev. 188, 1969 (1969).

[5] J. Ramsay, A.V. Gopal, E.M. Gauger, A. Nazir, B.W. Lovett, A.M. Fox, and M.S. Skolnick, Phys. Rev. Lett. 104, 017402 (2010).

[6] J.R. Schaibley, A.P. Burgers, G.A. McCracken, D.G. Steel, A.S. Bracker, D. Gammon, and L.J. Sham, Phys. Rev. B 87, 115311 (2013).

[7] C. Cohen-Tannoudji, J. Dupont-Roc and G. Grynberg, *Atom-Photon Interactions,* Wiley-VCH, Weinheim (2004).

[8] C. Roy and S. Hughes, Phys. Rev. B 85, 115309 (2012).

[9] D.P.S. McCutcheon and A. Nazir, Phys. Rev. Lett. 110, 217401 (2013).

[10] A.P. Saiko, G.G. Fedoruk, S.A. Markevich. Pis'ma v Zh. Eksp. Teor. Fiz. 98, 228 (2013).

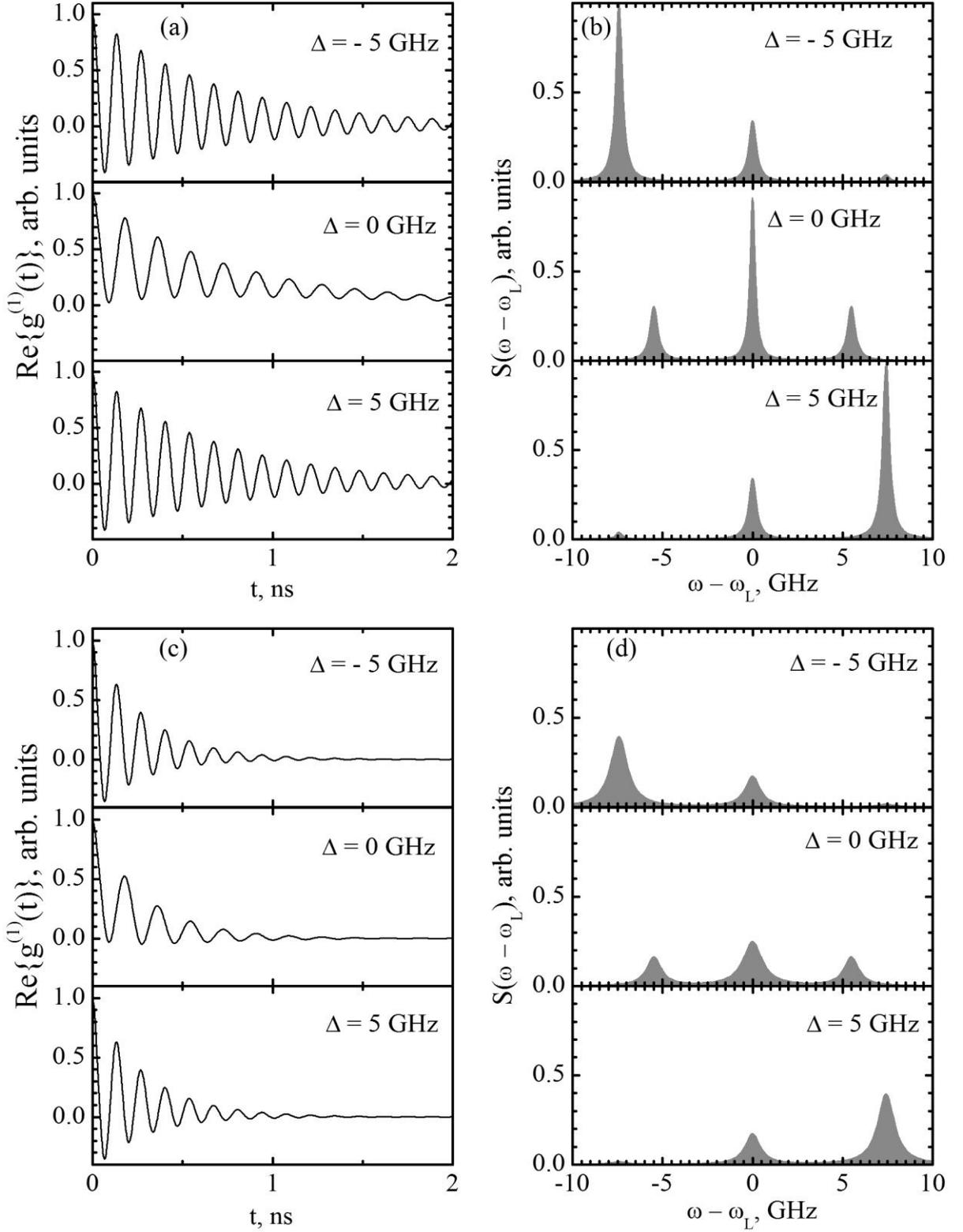

Fig. 1. (a), (c) The two-time correlation function and (b), (d) the QD emission spectrum for different values of laser-QD detuning at a fixed $\Omega/2\pi = 5.5$ GHz. The parameters employed are $\gamma_\parallel = 0.2$ GHz, $\gamma_{ph} = 0.12$ GHz, (a) and (b) for $\eta = 0$, (c) and (d) for $\eta = 0.94$ GHz.

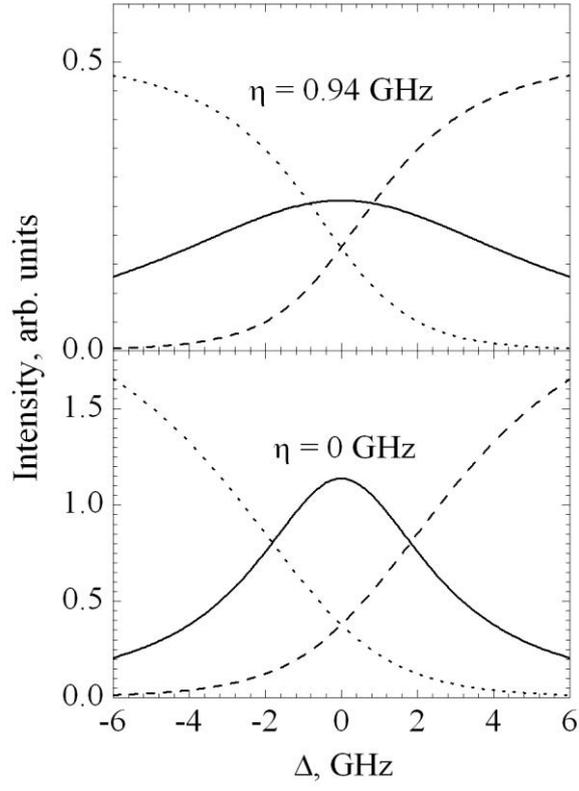

Fig. 2. The amplitude of the blue (dot line), red (dashed line) and central (solid line) peaks of the Mollow triplet versus laser-QD detuning at a fixed $\Omega/2\pi = 5.5$ GHz for (a) $\eta = 0$ and (b) $\eta = 0.94$ GHz. The relaxation rates $\gamma_\parallel$ and $\gamma_{ph}$ are the same as in Fig. 1.

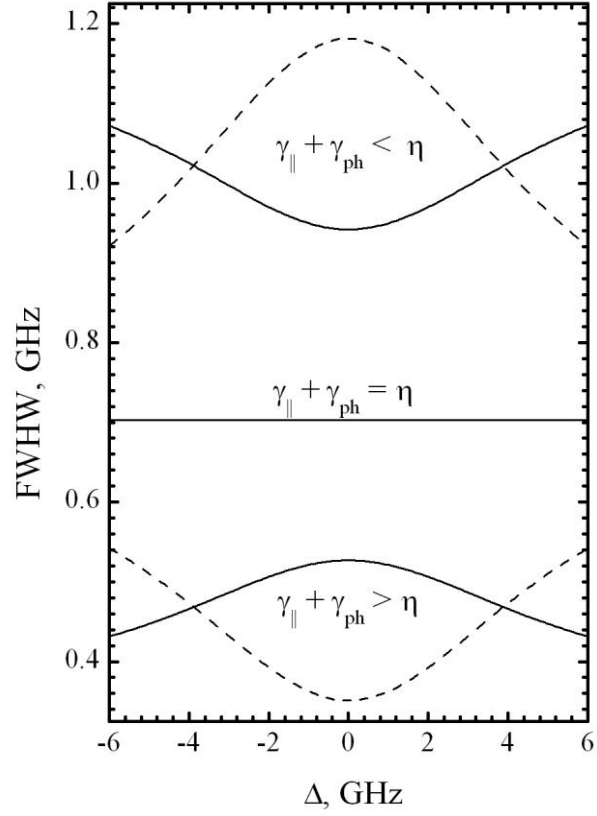

Fig. 3. The widths of the side (solid line) and central (dashed line) peaks of the Mollow triplet as a function of laser-QD detuning at a fixed $\Omega/2\pi = 5.5$ GHz for $\eta = 0$, $\eta = \gamma_{\parallel} + \gamma_{ph}$ and $\eta > \gamma_{\parallel} + \gamma_{ph}$ ($\eta = 0.94$ GHz). The relaxation rates $\gamma_{\parallel}$ and $\gamma_{ph}$ are the same as in Fig. 1.

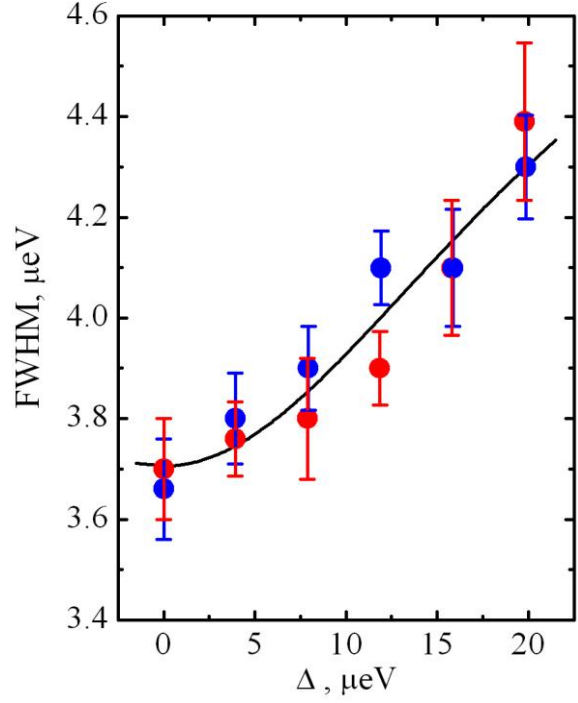

Fig. 4. (Color online) The FWHM of the Mollow sidebands versus laser-QD detuning. Circles denote the experimental results adopted from [3]. The solid line is calculated using equation (12) with the following parameters: $\Omega = 22.7$ μeV $\gamma_{\parallel} = 0.84$ μeV, $\gamma_{ph} = 0.34$ μeV, $\eta = 3.89$ μeV.

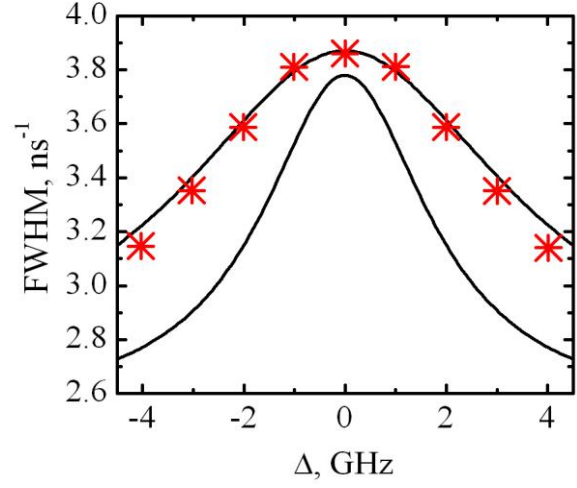

Fig. 5. The FWHM of the Mollow sidebands as a function of the laser-QD detuning at two fixed Rabi frequencies $\Omega/2\pi = 4$ GHz, $\gamma_{ph} = 0.08$ ns$^{-1}$ (the black line) and $\Omega/2\pi = 2$ GHz, $\gamma_{ph} = 0.02$ ns$^{-1}$ (the grey line) for $\eta = 0$ and $\gamma_{\parallel} = 2.5$ ns$^{-1}$. The symbols correspond to the numerical calculations of Ref. [9] for the red (×) and blue (+) sidebands.